\documentclass[reprint,amsmath,amssymb,citeautoscript,aip,jcp,superscriptaddress]{revtex4-1}

\usepackage{graphicx}
\usepackage{dcolumn}
\usepackage{bm}
\usepackage{xcolor}
\usepackage{hyperref}

\definecolor{darkpastelgreen}{rgb}{0.01, 0.75, 0.24}
\definecolor{emerald}{rgb}{0.31, 0.78, 0.47}

\newcommand{\tN}{t_{\mathrm{N}}}

\newcommand{\kB}{k_\mathrm{B}}
\newcommand{\Nw}{N_{\mathrm{w}}}

\newcommand{\ellB}{\ell_{\mathrm{B}}}

\newcommand{\ve}[1]{\mathbf{#1}}
\newcommand{\lambdaid}{\lambda_{\mathrm{id}}}

\begin{document}

\title{On the molecular correlations that result in field-dependent conductivities\\ in electrolyte solutions}
\author{Dominika Lesnicki}\thanks{These authors contributed equally}
\affiliation{Sorbonne Universit\'e, CNRS, Physico-Chimie des Electrolytes et
Nanosyst\`emes Interfaciaux,  Paris, France \looseness=-1}
\author{Chloe Y. Gao} \thanks{These authors contributed equally}
 \affiliation{Department of Chemistry, University of California, Berkeley, California \looseness=-1}
\author{David T. Limmer}
 \email{dlimmer@berkeley.edu}
 \affiliation{Department of Chemistry, University of California, Berkeley, California \looseness=-1}
\affiliation{Kavli Energy NanoScience Institute, Berkeley, California \looseness=-1}
\affiliation{Materials Science Division, Lawrence Berkeley National Laboratory, Berkeley, California \looseness=-1}
\affiliation{Chemical Science Division, Lawrence Berkeley National Laboratory, Berkeley, California \looseness=-1}
\author{Benjamin Rotenberg}
\affiliation{Sorbonne Universit\'e, CNRS, Physico-Chimie des Electrolytes et
Nanosyst\`emes Interfaciaux,  Paris, France \looseness=-1}
\affiliation{R\'eseau sur le Stockage Electrochimique de l'Energie (RS2E), FR CNRS 3459, France \looseness=-1}

\date{\today}

\begin{abstract}
Employing recent advances in response theory and nonequilibrium
ensemble reweighting, we study the dynamic and static correlations that give
rise to an electric field-dependent ionic conductivity in electrolyte solutions.
We consider solutions modeled with both implicit and explicit solvents, with
different dielectric properties, and at multiple concentrations. Implicit
solvent models at low concentrations and small dielectric constants exhibit
strongly field-dependent conductivities. We compared these results to the
Onsager-Wilson theory of the Wien effect, which provides a qualitatively
consistent prediction at low concentrations and high static dielectric
constants, but is inconsistent away from these regimes. The origin of the
discrepancy is found to be increased ion correlations under these conditions.
Explicit solvent effects act to suppress nonlinear responses, yielding a weakly
field-dependent conductivity over the range of physically realizable field
strengths. By decomposing the relevant time correlation functions, we find that
the insensitivity of the conductivity to the field results from the persistent
frictional forces on the ions from the solvent. Our findings illustrate the
utility of nonequilibrium response theory in rationalizing nonlinear transport behavior. 
\end{abstract}


\maketitle

\section{Introduction}
Due to their ubiquity and importance, electrolyte solutions have been central to
the development of theoretical physical chemistry.\cite{harned1959physical}
Early research into their structure and dynamics ushered in modern theoretical
techniques employing pair distribution functions and linear response
theories.\cite{zwanzig1970dielectric,onsager1934surface,onsager2002irreversible,kubo1956general}
Motivated by developments in the theory of systems far from equilibrium and
growing experimental capabilities for probing highly nonlinear transport
behavior, Gao and Limmer developed a theory of nonlinear response based on a
large deviation formalism.\cite{gao2018nonlinear}   Applying this theory to
ionic solutions in recent work\cite{lesnicki2020field}, we demonstrated the
ability to employ generalized response relations valid within nonequilibrium
steady states together with a nonequilibrium ensemble reweighting scheme to
discern the electric field-dependence of the ionic conductivity. In this paper,
we examine these relations further, considering models with both implicit and
explicit solvent, and comparing our findings to existing analytical theories
valid in dilute regimes.  We find generally that we are able to estimate the
field-dependent conductivity with high accuracy and that its form reflects an
interplay between ionic relaxational effects and solvent friction.

Efforts to predict the response of systems far from equilibrium from underlying
molecular properties have been stimulated recently in part by a growing body of
experimental observations on nanofluidic
devices.\cite{bocquet2010nanofluidics,bocquet_nanofluidics_2020,kavokine2021fluids}
In these systems, nonlinear transport is especially pervasive and often
qualitatively sensitive to chemical composition.  For example, ionic mobilities
have been observed to depend on the driving force of the flow, in a manner
dependent on the ion pair and the confining material.\cite{mouterde2019molecular}
Other nonlinear responses, such as ionic rectification, can be achieved in
nanofluidic diodes or conical pores.\cite{karnik2007rectification,
he2009tuning,picallo_nanofluidic_2013,cheng_nanofluidic_2010} Traditional linear
response theory such as Green-Kubo relations are incapable of describing such
behaviors due to the breaking of time-reversal symmetry within a steady-state
driven by a constant applied field.\cite{zwanzig1965time} Stochastic
thermodynamics have offered means of moving beyond equilibrium linear response
theory,\cite{seifert2012stochastic} and a number of generalized
fluctuation-dissipation relations and bounds have been proposed within its
context.\cite{speck2006restoring,baiesi2009fluctuations,prost2009generalized,dechant2020fluctuation}
Using large deviation theory,\cite{touchette2018introduction} a nonequilibrium
trajectory ensemble framework has been proposed in the context of field driven
steady-states for stochastic equations of motion that enables the relationship
between fluctuations of time extensive observables in or out of equilibrium to
be transparently derived.\cite{gao2017transport,gao2018nonlinear} Applications
to transport and self-assembly have been elucidated in simple systems.\cite{ray2019heat,lesnicki2020field,das2021variational,grandpre2021entropy}

\begin{figure}
\begin{center}
\includegraphics[width=8.5cm]{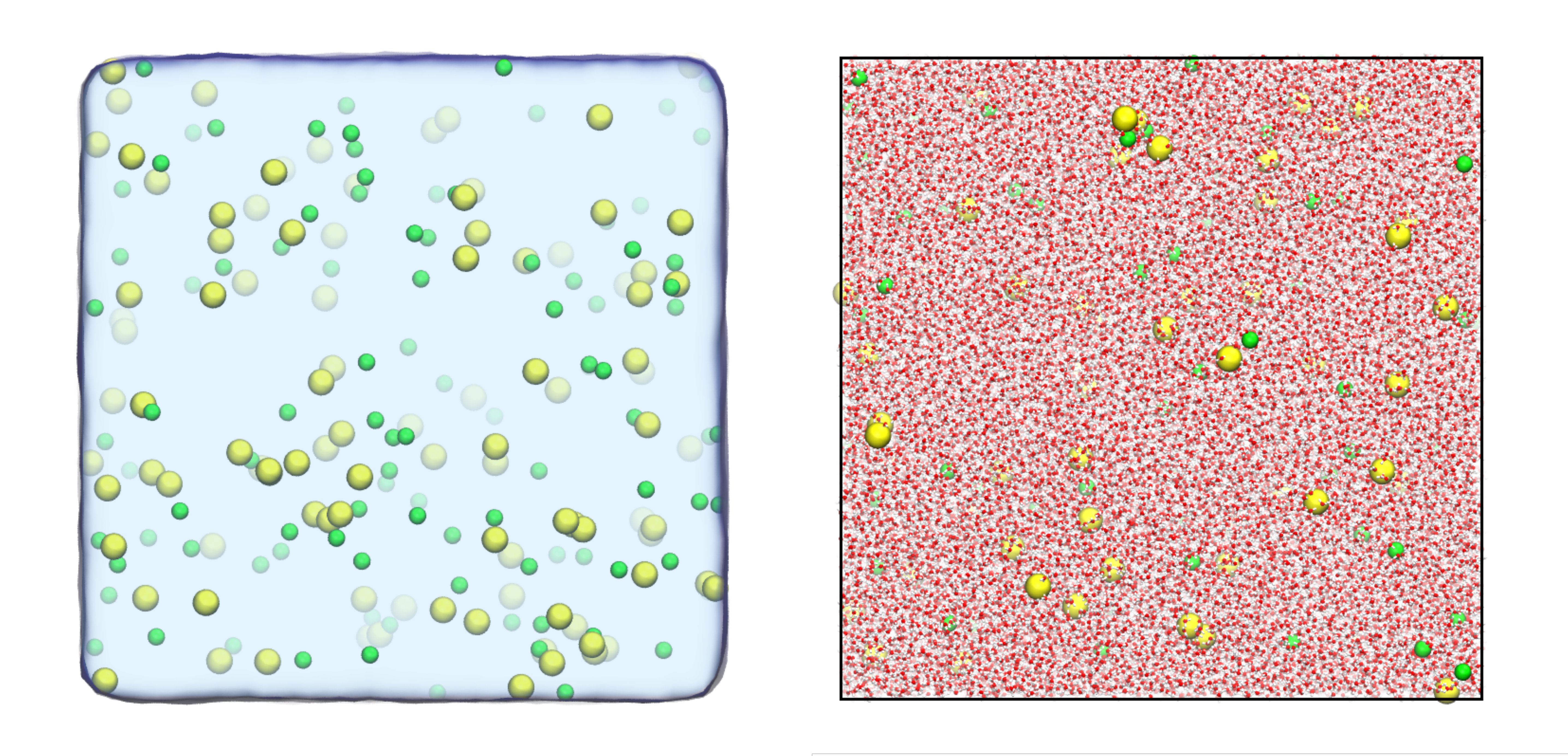}
\caption{\label{Fig0} Characteristic snapshots of NaCl solutions from the molecular dynamics simulations with implicit (left) and explicit (right) solvent. }
\end{center} 
\end{figure}

In this work, we extend these efforts to decode the relationship between
dynamical fluctuations and response far from equilibrium to systems with full
molecular detail in the context of ionic conductivities of electrolyte
solutions. To accomplish this, we devise methods of efficiently evaluating the
probability of rare dynamical fluctuations through a nonequilibrium ensemble
reweighting principle. With access to these probability distributions, we are
able to compute averaged observables like the differential conductance and ion
pair correlation functions for arbitrary applied fields and concentrations. In
what follows, we consider separately models of electrolyte solutions in which
the solvent is implicit, allowing us to compare to the Onsager-Wilson continuum theory\cite{wilson1936theory} of the Onsager-Wien effect,\cite{onsager1957wien} and those in which the solvent is explicit, allowing us to elucidate the role of ion-water correlations in the conductivity. Snapshots of the systems are shown in Fig.~\ref{Fig0}. In each section, the theory and numerical methods are first detailed followed by a discussion of the numerical results. 
 
\section{Implicit Solvent Model}
\label{sec:implicit}

We first consider a model under implicit solvent condition. Specifically, we
study a 0.1~M NaCl electrolyte composed of $N_{\mathrm{a}}$ anions and $N_{\mathrm{c}}$ cations with
$N=N_{\mathrm{a}}+N_{\mathrm{c}}$ in a volume $V$ and fixed temperature, $T$=300~K.
The particles' positions and velocities are denoted, $\ve{r}^N=\{ \ve{r}_1,\ve{r}_2, \dots,\ve{r}_N \}$ and $\ve{v}^N=\{\ve{v}_1,\ve{v}_2, \dots,\ve{v}_N\}$, respectively. These variables evolve according to an underdamped Langevin equation,
\begin{equation}
\label{Eq:Lang}
\dot{\ve{x}}_i=\ve{v}_i \, ,\quad m_i \dot{\ve{v}}_i = - \zeta_i \ve{v}_i + \ve{F}_i \left (\ve{r}^N \right) + z_i \ve{E}  + \bm{\eta}_i
\end{equation}
where $m_i$ and $z_i$ are the $i$th particle's mass and charge, $\zeta_i$ is the
friction and $\ve{F}_i(\ve{r}^N)$ is the
interparticle force on the particle $i$. The ions interact through a screened
 Coulomb potential with dielectric constants
$\varepsilon$ = 10 or 60, and Lennard-Jones (LJ) potential. The solutions with low and high values of the dielectric constant we refer to as weak and strong electrolytes, respectively, referring to whether the screening length is comparable or larger than the characteristic ionic size.
For pairs $ij$ the total potential $U_{ij}(r)$ is
\begin{equation}
\label{EqLJ}
U_{ij}(r) = 4\epsilon_{ij} \left[ \left( \frac{\sigma_{ij}}{r}\right)^{12} - \left( \frac{\sigma_{ij}}{r}\right)^6 \right] + \frac{1}{4 \pi \varepsilon_0 \varepsilon} \frac{z_i z_j}{r}
\end{equation}
where $\sigma_{ij}$ and $\epsilon_{ij}$ are the LJ length and energy parameters.
To model Na$^+$ and Cl$^-$, we adopt the parameters of Ref.~\citenum{Koneshan1998NaCl}. The usual Lorenz-Berthelot combining rules, $\sigma_{ij} = (\sigma_i + \sigma_j)/2$ and $\epsilon_{ij} = (\epsilon_i \epsilon_j)^{1/2}$, were used to calculate the LJ interactions.  Each cartesian component of the random force, ${\eta}_{i\alpha}$, obeys
Gaussian statistics with mean $\langle {\eta}_{i\alpha} \rangle =0$ and
variance $\langle {\eta}_{i\alpha}(t){\eta}_{j\beta}(t') \rangle = 2 \kB
T \zeta_i \delta_{ij} \delta_{\alpha\beta} \delta(t-t')$, where $\kB$ is
Boltzmann's constant. Finally, $\ve{E}$ denotes an applied electric field, with magnitude $E$ that drives an ionic current through the periodically replicated system.\\

In order to reproduce the dynamics of the ions in explicit solvent (see
Sec.~\ref{sec:explicit}), we use the dielectric constant $\varepsilon =60$ of
the corresponding water model and the frictions corresponding to the
diffusion coefficients calculated for the ions at infinite dilution in the
explicit solution: $\zeta_i = m_i/\tau_i$ with
relaxation times $\tau_{\mathrm{c}} = 10.1$~fs and $\tau_{\mathrm{a}} = 21.3$~fs, for the cations
and anions with masses $m_{\mathrm{c}}= 22.99$~a.m.u. and
$m_{\mathrm{a}}=35.45$~a.m.u., respectively. 
These large frictions effectively
render the dynamics overdamped, and we explicitly neglect hydrodynamic effects
resulting from integrating out the surrounding solvent
flow.\cite{jardat1999transport,ladiges2021discrete}
Simulations are performed for 100 ion pairs, using the LAMMPS
code\cite{plimpton1995a,LAMMPScode} with a modified Langevin thermostat to
ensure a Gaussian distribution of the noise. 
Results for the conductivity are obtained from 10~ns nonequilibrium
simulations for finite fields between 0 and 0.1~V/\AA, in steps of 0.01~V/\AA, with statistical error estimated from bootstrapping,
while those for spatial correlations are obtained from 10 independent 10~ns trajectories.

\subsection{Nonlinear response from trajectory reweighting}

To compute the molar ionic conductivity as a function of the electric field, we
employ the formalism of our previous work based on the reweighting of
nonequilibrium trajectories.~\cite{lesnicki2020field} 
The differential conductivity $\sigma(E)$ and its molar counterpart $\lambda(E)$
are defined from the derivative of the current with respect to the field,
\begin{equation}\label{def}
\lambda(E) = \frac{1}{N} V\sigma(E) = \frac{1}{N} \frac{d\langle J\rangle_E}{dE}
\; ,
\end{equation}
where given the trajectory, $\ve{X}(\tN)$, or sequence of positions and velocities 
over an observation time, $\tN$, the current is given by
\begin{equation}
\label{Eq:Current}
J[\ve{X}(\tN)] = \frac{1}{\tN}  \int_0^{\tN} dt \, j(t) \, ,\quad j(t)= \sum_{i=1}^N z_i v_i(t) 
\end{equation}
a time average of a charge weighted ionic velocity. The $\langle \dots
\rangle_E$ denotes an average over a trajectory ensemble, where the ions are
evolved under the action of an electric field with magnitude $E$. Assuming the
fluid is isotropic, the velocity in Eq. ~\ref{Eq:Current} is the component in
the direction of the applied field.
Note that the current $J$ is usually defined in experiments with a factor
$1/V$, which explains that with the present notations $d\langle J\rangle_E/dE=V\sigma(E)$.

In order to compute Eq.~\ref{def} efficiently, we relate the current of a system perturbed by an additional applied field to a system with no applied field. Given the equation of motion in Eq.~\ref{Eq:Lang}, the probability of observing a trajectory, $\ve{X}(\tN)$  with an applied field, is
\begin{equation}
\label{Eq:Measures}
P_{\ve{E}}[\ve{X}(\tN)] \propto e^{-\beta U_{\ve{E}}[\ve{X}(\tN)]}
\end{equation}
where $\beta=1/k_BT$ and for the uncorrelated Gaussian noise, we have an Onsager-Machlup stochastic action of the form\cite{kleinert2009path}
\begin{align}
U_{\ve{E}}[\ve{X}(\tN)]& =
\sum_{i=1}^N  \int_0^{\tN}  dt \frac{\left [m_i \dot{\ve{v}}_i + \zeta_i
	\ve{v}_i - \ve{F}_i(\ve{r}^N) -z_i  \ve{E}  \right ]^2}{4 \zeta_i}  \,   
\end{align}
where the stochastic calculus is interpreted in the It\^o sense.  We will
consider trajectories in the limit that $\tN$ is large so that only time extensive quantities are relevant. 

A perturbing field on the system adds an extra drift to the Gaussian action. As a consequence, we can write the ratio
of the probability to observe a trajectory in the presence of the field, denoted as
$\ve{E}$, relative to no applied field through a Girsanov transform\cite{girsanov1960transforming}
\begin{equation}
\label{Eq:Measures2}
\frac{P_{\ve{E}}[\ve{X}(\tN)]}{P_{\ve{0}}[\ve{X}(\tN)]} = e^{\beta \Delta U_{\ve{E}}[\ve{X}(\tN)]}
\end{equation}
where the dimensionless relative action, $\beta \Delta U_{\ve{E}}[\ve{X}(\tN)]$, can be expressed compactly as a sum of three terms, depending on their symmetry under time reversal\cite{basu2015nonequilibrium}
\begin{align}
\label{Eq:DeltaUDeltaE}
\frac{\Delta U_{\ve{E}}}{\tN} = 
 J \frac{E}{2} +  Q \frac{E}{2} 
- N \lambdaid \frac{E^2}{4}  \, 
\end{align}
where for simplicity we take the field along one cartesian direction so that the relative action depends only on its magnitude.
The first term is asymmetric under time reversal and identified as the excess
entropy production due to the increased nonequilibrium driving. It is given by
the product of the total, time averaged current in the direction of the field and the field $E$.
The second term in Eq.~\ref{Eq:DeltaUDeltaE} is symmetric under time reversal
and is the product of the field by a quantity referred to as the excess frenesy\cite{baiesi2009fluctuations} 
\begin{eqnarray}
\label{Eq:Fenesy}
Q[\ve{X}(\tN)] &=& \frac{1}{\tN}  \int_0^{\tN} dt \, q(t)  \\
{\rm with}\ q(t) &=& \sum_{i=1}^N \frac{z_i}{\zeta_i} \left [ m_i\dot{v}_i(t)- F_i(\ve{r}^N)\right ] \nonumber
\; ,
\end{eqnarray}
which includes the total time integrated force in the direction of the field weighted
by $z_i/\zeta_i$, and a boundary term resulting in a difference in velocities at
times 0 and $\tN$. The remaining term in Eq.~\ref{Eq:DeltaUDeltaE} is a trajectory independent constant,
\begin{equation}
\label{Eq:Sig0}
\lambdaid  = \sum_{i=\mathrm{a},\mathrm{c}}  \frac{N_i}{N} \lambda_i 
= \sum_{i=\mathrm{a},\mathrm{c}}  \frac{N_i}{N} \frac{D_i z_a^2}{\kB T} 
\end{equation}
where $D_i = \kB T/\zeta_i$ is the diffusion coefficient for an isolated ion of
type $i$. This constant is the molar Nernst-Einstein conductivity of the
solution and is fully determined by the nature of the electrolyte under finite dilutation, \emph{i.e.}
the molar fraction, charge and diffusion coefficients of the independent ions. 

With the relative measure between trajectory ensembles defined in
Eq.~\ref{Eq:Measures2}, we can relate nonequilibrium trajectory averages in
the presence of the field, to equilibrium trajectory averages
without the field. For a trajectory observable, $O[\ve{X}(\tN)]$, this relation is
\begin{equation}
\label{Eq:ReweightO}
\left \langle O \right \rangle_E =\left \langle O e^{\beta \Delta U_{\ve{E}}[\ve{X}(\tN)]} \right \rangle_0
\end{equation}
where trajectory averages are over the measure in Eq.~\ref{Eq:Measures} with fields $E$ and 0. Setting $O$ to 1, we find a sum rule inherited from the underlying Gaussian process that is quadratic in the field,
\begin{equation}
\left \langle e^{\beta \tN (J[\ve{X}(\tN)] + Q[\ve{X}(\tN)] )E /2} \right
\rangle_0 = e^{\beta \tN  N \lambdaid  E^2/4}
\end{equation}
which is interpretable as the ratio of nonequilibrium to equilibrium trajectory partition functions.

Identifying the joint probability of observing a value of the current and frenesy as $p_E(J,Q)=\langle \delta(J-J[\ve{X}(\tN)],Q-Q[\ve{X}(\tN)]) \rangle_E$, we can relate $p_E(J,Q)$ to its equilibrium counterpart, using Eq.~\ref{Eq:ReweightO},
\begin{equation}
\label{Eq:ReweightP}
\frac{\ln p_0(J,Q)}{\tN} = \frac{\ln p_E(J,Q)}{\tN} - \beta (J +Q) \frac{E}{2}
+ \beta N \lambdaid \frac{E^2}{4}
\end{equation}
and thus have a means by which sampling simulations at a variety of finite fields, we are able to reconstruct the joint distribution, $p_0(J,Q)$, far into its tails in a manner similar to parallel tempering or replica exchange.\cite{earl2005parallel,sugita2000multidimensional} Moreover, with knowledge of the full $p_0(J,Q)$, we can in principle compute the molar conductivity $\lambda (E)$ as a continuous function of the applied field. This is done by substituting Eq.~\ref{def} into Eq.~\ref{Eq:ReweightO},
\begin{equation}
\label{Eq:Reweightlam0}
\lambda (E)=
\lim_{t_{\textrm{N}}\rightarrow\infty}\frac{\beta t_{\textrm{N}}}{2N}\left\langle (\delta J^2+\delta J\delta Q)e^{\beta \Delta U_{\ve{E}}(J,Q)}\right\rangle_0
\end{equation}
which provides a direct means of evaluating the conductivity. 
This relation extends the more familiar Einstein-Helfand
expression\cite{helfand_transport_1960,zwanzig2001nonequilibrium}, which is
recovered in the absence of the external field.\cite{lesnicki2020field}
As we have detailed previously, this route is statistically superior
to the explicit estimate afforded from a finite difference approximation to the
derivative of the current with applied field. In addition, the above estimator
provides information on the conductivity away from points of explicit simulation
that is reliable as along as the trajectories ensembles are well
overlapping.\cite{frenkel2001understanding} We note that as in equilibrium, fluctuation relations like that above can be found for differential quantities, unlike their integral counterparts.\cite{limmer2013charge}


\subsection{Comparison to Onsager-Wilson theory}

\begin{figure}[h!]
\begin{center}
\includegraphics[width=8.5cm]{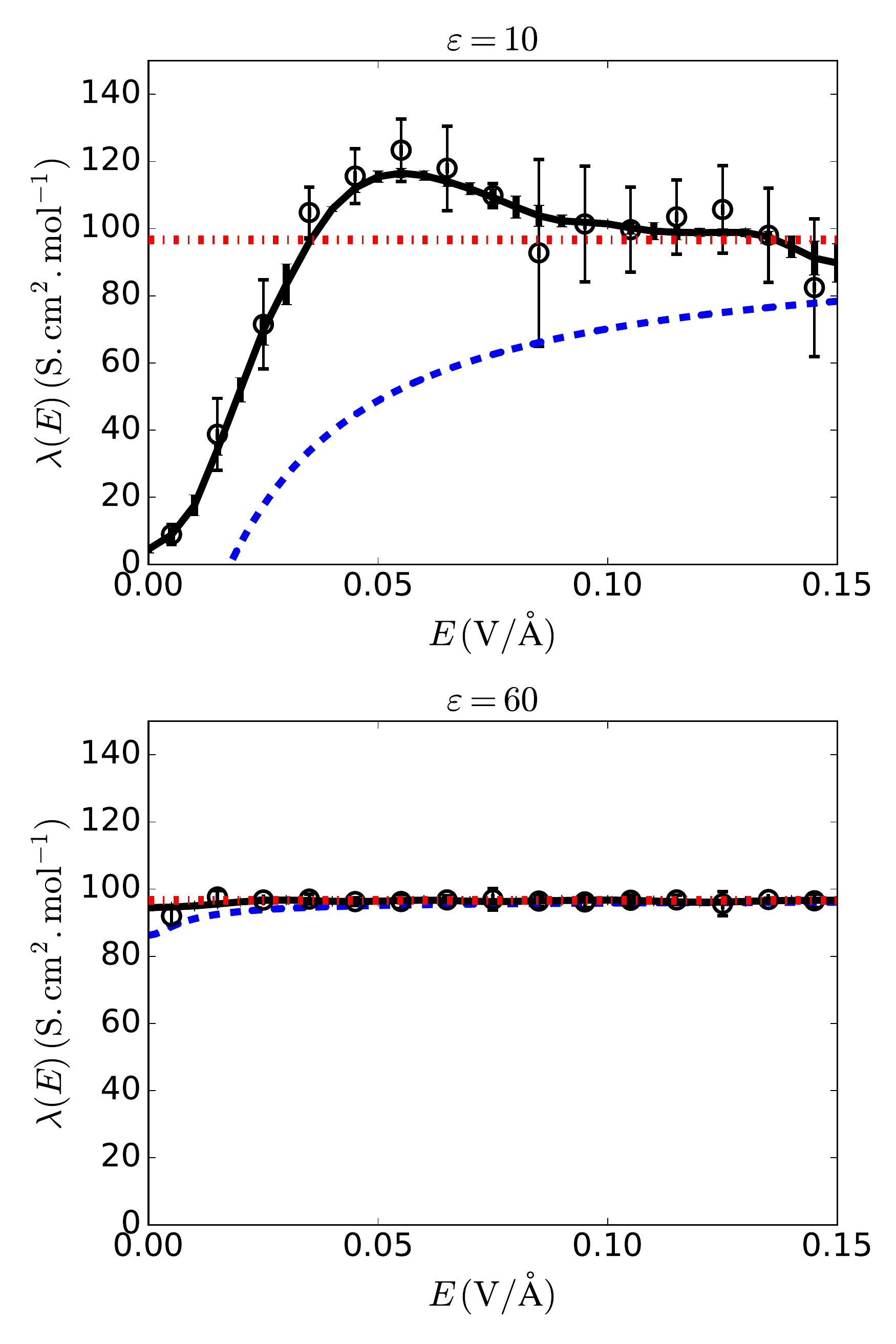}
\caption{\label{Fig1}Field-dependent molar ionic conductivities for $\varepsilon = 10$ (top panel) and 60 (bottom panel). Black lines
are computed from reweighting $p(J, Q)$, and symbols are computed from finite
differences of $\left<J\right>_E$ versus $E$. Errorbars are one standard
deviation of the mean. The red dashed lines correspond to the Nernst-Einstein
molar ionic conductivities $\lambdaid$ (Eq. \ref{Eq:Sig0}). The blue dashed
lines correspond to the Wilson molar ionic conductivities considering solely the
effect of drag due to the ionic atmosphere (see Eq.~\ref{Eq:Wilson}). }
\end{center} 
\end{figure}

Shown in Fig.~\ref{Fig1} are the molar ionic conductivities computed from Eq.~\ref{Eq:Reweightlam0} continuously as a function of the applied field and from
a numerical derivative of the average current versus applied field smoothed with
a spline function. In order to apply Eq.~\ref{Eq:Reweightlam0}, we constructed
$p_0(J, Q)$ with a series of simulations at finite fields at the locations of
the direct estimate in Fig.~\ref{Fig1}, using a generalized version of
WHAM\cite{kumar1992weighted} employing the ensemble relation in
Eq.~\ref{Eq:ReweightP} to stitch joint histograms of $J$ and $Q$ together.  We
find quantitative agreement between both estimates, but the calculation
employing the ensemble reweighting principle has systematically smaller
statistical errors. This is due in part to the avoidance of taking a numerical
derivative and in part due to the ability to use information on the conductivity
from adjacent fields. 

As previously observed,\cite{lesnicki2020field} for the weak electrolyte
$\varepsilon = 10$, we find a field-dependent conductivity that increases
initially quadratically before plateauing at large fields to the Nernst-Einstein
limit. At intermediate fields, the weak electrolyte exhibits a slight maxima in
ionic molar conductivity. This peak is a consequence of markedly non-Gaussian
current fluctuations, with fat tails that are emphasized by the nonequilibrium
reweighting at finite field. This behavior is in contrast to that of the strong
electrolyte $\varepsilon = 60$, which exhibits a nearly field-independent
conductivity. The strong electrolyte conductivity is found to saturate to the
Nernst-Einstein limiting value for the smallest values of the applied fields
considered. 

The field-dependent conductivity was considered theoretically by
Wilson,\cite{wilson1936theory} following foundational work on the linear
conductivity by Onsager and Fuoss.\cite{onsager2002irreversible} In his work,
Wilson considered a strong electrolyte in the dilute limit, in which the pair
correlations are well described by Debye-H\"uckel theory in the absence of an
applied field. 
\footnote{For historical context concerning the work of Wilson, reviewed extensively in Ref \onlinecite{harned1959physical} but published only in his PhD thesis, see Ref ~\onlinecite{eu2010onsager}.}
From the field-dependence of the pair correlation
functions, a force balance on the ions leads to the following expression of
the molar conductivity\cite{lesnicki2020field} 
\begin{equation}
\lambda(E) = \lambdaid +\sum_{i,j=\mathrm{a},\mathrm{c}}\frac{z_i}{\zeta_i}\int
{\rm d}\ve{r} \, \bar{\rho}_j g'_{ij}(r,\theta|E) F_{ij,x}(\ve{r})
\end{equation}
where $r=|\ve{r}|$, $\cos(\theta) = \ve{r}\cdot \hat{\ve{x}}$ and
$g'_{ij}(r,\theta|E) = dg_E(r,\theta)/dE$ is the derivative of the $ij$ pair
distribution function at finite field $E$. The force $F_{ij,x}(\ve{r})$ is the
component of the pair force in the direction of the field, here assumed to be
along the $x$ direction and $\bar{\rho}_j=N_j/V$. 

The Onsager-Wilson predictions are shown in Fig.~\ref{Fig1}, with the complete
expressions for $\lambda(E)$  shown in App.~\ref{Sec:Wilson}. In
Fig.~\ref{Fig1}, we have included only the term from ionic relaxation, not the
additional electrophoretic terms that result from the counterflow of the solvent
(\emph{i.e.} $\Delta\sigma_{\rm rel}$ but not $\Delta\sigma_{\rm hyd}$ in
Eq.~\ref{Eq:Wilson}).  This is because in our implicit model, the latter
effect is not present. Comparing the theoretical prediction to our numerical
results we find poor agreement for the low dielectric system, with Wilson's
theory predicting an unphysical negative conductivity before increasing towards
the Nernst-Einstein limit. This failure is due to the break down of the
Debye-H\"uckel approximation, as the low dielectric constant results in a large, nonperturbative interaction between the ions. The agreement is better for the high dielectric
system, though the theory predicts a smaller conductivity at zero applied field
and a larger response to the field than observed in the simulation.

\begin{figure}[t]
	\begin{center}
		\includegraphics[width=8.5cm]{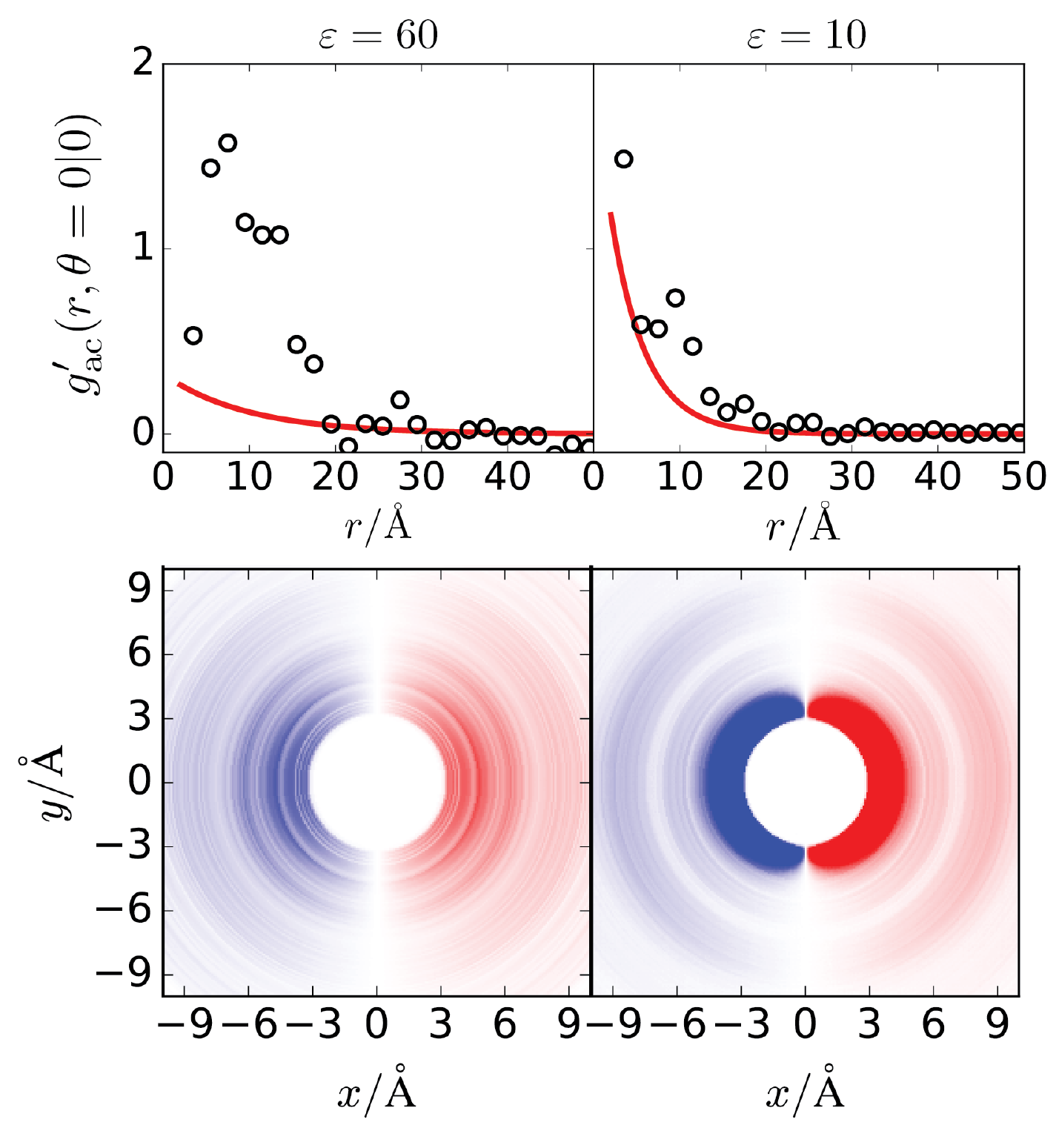}
		\caption{\label{fig:MapOnsager}
Response of the pair distribution to an applied field for
cations around an anion, at equilibrium ($E=0$) for $\varepsilon = 60$
(left) and $\varepsilon = 10$ (right).
Results are shown (top) for Onsager-Wilson theory Eq.~\ref{Eq:G_Onsager} in the solid red line and from simulations using Eq.~\ref{Eq:GrE} in the black symbols. Response function, $g_{ac}'(x,y|0)$,
 in the $(x,y)=(r\cos\theta,r\sin\theta)$ plane (bottom)
 from simulations using Eq.~\ref{Eq:GrE}. The red (blue) color indicates 
excess (depletion) of cations in the presence of a field oriented from
left to right. The color scale is conserved in both panels but saturates at
lower and higher extremes.
}
	\end{center} 
\end{figure}

In order to understand the relationship between our numerical results and
Onsager-Wilson theory, we compare the response functions of the pair correlation
functions directly. The pair distribution, $g_{\mathrm{a}\mathrm{c}}(r, \theta|E)$, between
an anion and a cation can be computed exactly, up to quadrature~\cite{Onsager_collectedworks,eu2010onsager}, and its
first order correction with applied field is simple. For a symmetric electrolyte,
with charges $z=\pm qe$, Onsager's theory predicts for cations around anions
\begin{eqnarray}\label{Eq:G_Onsager}
	g'_{\mathrm{a}\mathrm{c}}(r, \theta| 0) =&& - \cos(\theta) \frac{\beta^2 z^3}{2 \pi \varepsilon_0 \varepsilon \kappa^2 r^2} \times \\
		 && \left [ \left( 1+\kappa r \right ) e^{-\kappa r} -  \left( 1+\frac{\kappa r}{\sqrt{2}} \right ) e^{-\kappa r/\sqrt{2}} \right ] \nonumber
\end{eqnarray}
where $\kappa = \sqrt{4\pi \ellB \sum_{i=\mathrm{a},\mathrm{c}} \bar{\rho}_i q_i^2}$ is the inverse Debye screening length and $\ellB=\beta e^2/4 \pi \varepsilon_0 \varepsilon$ is the Bjerrum length. 
The predictions of Onsager theory for the response of the pair distribution functions are
shown in Fig.~\ref{fig:MapOnsager} for 0.1M, for both  $\varepsilon = 60$
($\kappa^{-1} = 8.4~\mathrm{\AA}$) and  $\varepsilon = 10$ ($\kappa^{-1}= 3.4~\mathrm{\AA}$). 
The response function is asymmetric, reflecting the polarization of the ionic
distribution due to the field. 
Specifically, there is a build up of cations around anions in
the direction of the field and a suppression in the opposite direction.
The radial extent of the response is much larger for $\varepsilon = 60$
than for $\varepsilon = 10$, reflecting the much stronger interactions between
the ions (hence smaller $\kappa^{-1}$) in the latter case. 

The calculation of the response of the ion pair correlation function to an applied electric field within our molecular dynamics simulations is complicated by the use of periodic boundary conditions and finite simulation volumes. The direct calculation from a finite difference of the correlation function evaluated in equilibrium and at small applied field predicts a polarization of the ion distribution erroneously opposite of that predicted by Onsager-Wilson theory. Instead of computing it from an explicit nonequilibrium simulation, we employ a linearized version of Eq.~\ref{Eq:ReweightO} to compute the response function directly within an equilibrium calculation. 

Denoting the total configurational distribution at time $\tN$,
$\rho_E(\ve{r}^N)=\langle \delta[\ve{r}^N-\ve{r}^N(\tN)]\rangle_E$, 
its form under applied field $E$ relative to its form in equilibrium is given by
\begin{equation}
 \rho_E(\ve{r}^N) = \rho_0(\ve{r}^N)  \frac{\langle e^{\beta E \tN  (J[\mathbf{X}(\tN)]+Q[\mathbf{X}(\tN)])/2} \rangle_{0,\ve{r}^N(\tN)}}{\langle  e^{\beta E \tN  (J[\mathbf{X}(\tN)]+Q[\mathbf{X}(\tN)])/2} \rangle_0} 
\end{equation}
where we have rewritten the numerator as an equilibrium average conditioned on ending in a particular configuration $\ve{r}^N(\tN)$ at time $\tN$. This form of a nonequilibrium distribution is similar to the Kawasaki distribution.\cite{yamada1967nonlinear,morriss1985isothermal,crooks2000path} 

Assuming that the field is small, we can linearize the exponentials to find the first order correction to the configurational distribution. Invoking time-reversal symmetry to move the conditioning to the initial time and a sum rule to eliminate $Q$,\cite{gao2018nonlinear} we arrive at,
\begin{equation}
 \rho_E(\ve{r}^N) =  \rho_0(\ve{r}^N) \left [ 1 -\beta E \tN \langle \delta J \rangle_{0,\ve{r}^N(0)} \right ] +\mathcal{O}(E^2)
\end{equation}
where $\langle \delta J \rangle_{0,\ve{r}^N(0)}=\langle  J \rangle_{0,\ve{r}^N(0)}-\langle  J \rangle_{0}$ is the difference in the time integrated current in the conditioned ensemble with its equilibrium value. This result demonstrates that the correction to the steady state distribution is just the entropy produced upon relaxation from a fixed initial configuration $\ve{r}^N_0$. The average current in equilibrium is zero, but subtracting it out explicitly produces a better estimator for the response function. Specializing to the pair distribution by integrating over all but two ion positions, for one ion of type $i$ and another of type $j$, 
\begin{equation}
\label{Eq:GrE}
 g'_{ij}(r,\theta | 0) = \lim_{t_{\textrm{N}}\rightarrow\infty}  -\beta  g_{ij}(r,\theta |0) \tN \langle \delta J \rangle_{0,r_{ij}(0),\theta_{ij}(0)} 
\end{equation}
we arrive at a numerically tractable means of computing the change of the ion
distribution function from an equilibrium simulation. Namely we correlate an
initial density to its transient current. Analogous results have been arrived at
through alternative means for computing the Green's function for the velocity
field due to point force in a fluid\cite{lesnicki2017microscopic} or the
mobility profile for a confined fluid.\cite{mangaud2020sampling} 

We find qualitative similarities to Onsager-Wilson theory, including an accumulation
of cations in front of the anion and a depletion at the back of the anion as
shown in Fig.~\ref{fig:MapOnsager}.  However, the magnitudes and
characteristic lengthscales are not quantitatively reproduced by the approximate
theory, compared in
Fig.~\ref{fig:MapOnsager}. For both cases, the theory does not
correctly predict the magnitude of the response. 
This is due to the weak coupling approximation used in Onsager-Wilson theory
that is valid only in the limit of small $\kappa$ and weak ion correlations. This is manifested in the qualitatively inaccurate
prediction of the field-dependent conductivity for the weak electrolyte, and a
stronger field-dependence than observed in the strong electrolyte. In the low
dielectric solution, ions interact strongly and
are much more responsive to the field in both the simulation and the theory. The ionic relaxation effect decreases the conductivity from the Nernst-Einstein value, and its overprediction is the cause of the spurious negative conductivity predicted in Fig. \ref{Fig1}.

\section{Explicit Solvent Model}
\label{sec:explicit}

We next consider a model with explicit solvent conditions. Specifically, we
study a NaCl electrolyte composed of $N_{\mathrm{a}}$ anions, $N_{\mathrm{c}}$ cations
and $N_{\mathrm{w}}$ water molecules in a volume $V$ and fixed temperature $T=300$K. We
will consider two concentrations of ions, 0.1 M and 1.0 M.  As before, the
species' positions and velocities are denoted, $\ve{r}^N=\{ \ve{r}_1,\ve{r}_2,
\dots,\ve{r}_N \}$ and $\ve{v}^N=\{\ve{v}_1,\ve{v}_2, \dots,\ve{v}_N\}$,
respectively. These variables evolve according to the same underdamped Langevin
equation as in Eq.~\ref{Eq:Lang}. The relaxation time $\tau=1000$ fs is used for
all species. The pairwise interaction potential $U_{ij}(r)$ consists of LJ and electrostatic terms such that
\begin{equation}\label{EqLJ}
U_{ij}(r) = 4\epsilon_{ij} \left[ \left( \frac{\sigma_{ij}}{r}\right)^{12} - \left( \frac{\sigma_{ij}}{r}\right)^6 \right] + \frac{1}{4 \pi \varepsilon_0} \frac{z_i z_j}{r}
\end{equation}
where $z_i$ and $z_j$ are the charges on sites $i$ and $j$, $r_{ij}$ is the site-site separation, $\sigma_{ij}$ and $\epsilon_{ij}$ are the length and energy parameters, and $\epsilon_0$ is the permittivity of free space.
The LJ parameters for Na$^+$ and Cl$^-$ are chosen the same as in the implicit
model, while the water molecules are modeled with the flexible q-TIP4P/F
model~\cite{habershon2009}. The usual Lorenz-Berthelot combining rules,
$\sigma_{ij} = (\sigma_i + \sigma_j)/2$ and $\epsilon_{ij} = (\epsilon_i
\epsilon_j)^{1/2}$, were used to calculate the LJ interactions. 
Simulations are performed for 100 ion pairs and 55508 (5540) water
molecules for the 0.1M (1M) solution, using the LAMMPS
code\cite{plimpton1995a,LAMMPScode} with a modified Langevin thermostat to
ensure a Gaussian distribution of the noise.
Results for the conductivity are obtained from a series of 1~ns nonequilibrium
simulations for finite field between 0 and 0.1~V/\AA, in steps of 0.01~V/\AA, with statistical error estimated from bootstrapping,
while those for temporal correlations are obtained from 10 independent 10~ns trajectories.

\subsection{Trajectory reweighting of multicomponent systems}

As in the implicit model, the stochastic action can be constructed by comparing the ratio of the probability to observe a trajectory with and without a perturbing field. The flexible water model employed allows for independent noises to act on each atom in the water molecule, while a rigid model would color the noise due to the imposed geometric constraint. 
For generality and for utility in subsequent analysis, we consider different
perturbing fields with magnitudes,
$\mathbf{E}=\{{E}_{\textrm{i}},{E}_{\textrm{w}}\}$ for the ions and water,
respectively. As before,  we take the field along one cartesian direction so
that the relative action depends only on these magnitudes. We further introduce the notation
$\mathbf{E^{(2)}}=\{{E}_{\textrm{i}}^2,{E}_{\textrm{w}}^2\}$. The stochastic action, decomposed according to symmetry under time reversal, is then
\begin{equation}
\frac{\Delta U_{\mathbf{E}}}{t_N}=\frac{1}{2} [\mathbf{J}+\mathbf{Q}] \cdot
\mathbf{E}-\frac{1}{4}  \boldsymbol{N\lambda} \cdot \mathbf{E}^{(2)}
\end{equation}
which now depends on components of the currents, $\mathbf{J}=\{J_{\textrm{i}},
J_{\textrm{w}}\}$, and excess frenesy, $\mathbf{Q}=\{Q_{\textrm{i}}, Q_{\textrm{w}}\}$, 
and vector $\boldsymbol{N\lambda}=\{ N_{\mathrm{a}}\lambda_{\mathrm{a}}+N_{\mathrm{c}}\lambda_{\mathrm{c}}, N_{\mathrm{w}}\lambda_{\mathrm{w}}\}$
from the ions and the water, respectively. 
The definitions of the dynamical variables are the same as in the implicit model, where sums are interpreted atom-wise over the specified group. Specifically for group $\alpha =\{\mathrm{i},\mathrm{w}\}$, the associated current is
\begin{equation}
\begin{aligned}
J_{\alpha}[\ve{X}(\tN)] &= \frac{1}{\tN}  \int_0^{\tN} dt \, j_{\alpha}(t) \,,\quad j_{\alpha}(t)=\sum_{i \in N_\alpha} z_i v_i(t)\\
\end{aligned}
\end{equation}
and frenesy, 
\begin{eqnarray}
Q_\alpha[\ve{X}(\tN)] &=& \frac{1}{\tN}  \int_0^{\tN} dt \, q_\alpha(t)  \\
 q_\alpha(t) &=& \sum_{i\in N_\alpha} \frac{z_i}{\zeta_i} \left [ m_i\dot{v}_i(t)- F_i(\ve{r}^N)\right ] \nonumber
\end{eqnarray}
The constant $\lambda_{\mathrm{w}}$ introduced above in $\boldsymbol{N\lambda}$ is defined for ions as in Eq.~\ref{Eq:Sig0}, with a sum over oxygen and hydrogen atoms,
but it no longer has the interpretation of a
contribution to the molar Nernst-Einstein conductivity. While the waters are flexible, they are
not dissociable, and their motion is described by a center of mass motion, with
additional rotational and vibrational modes. The charge neutrality of the water
molecule dictates that its center of mass motion does not contribute to the ionic
current, however rotational and vibrational motions may, transiently. 

With the relative measure between trajectory ensembles defined in
Eq.~\ref{Eq:Measures2}, we can relate nonequilibrium trajectory averages in
the presence of the field, to equilibrium trajectory averages
without the field. Following our previous work\cite{lesnicki2020field}, an average of an arbitrary observable $\mathcal{O}$ under an applied field can be expressed as
\begin{equation}
\label{Eq:Current_Explicit}
\left\langle \mathcal{O}\right\rangle_E=\frac{\left \langle \mathcal{O} e^{\frac{\beta \tN}{2} (\mathbf{J}+\mathbf{Q}) \cdot \mathbf{E}} \right \rangle_0}{\left \langle  e^{\frac{\beta \tN}{2} (\mathbf{J}+\mathbf{Q}) \cdot \mathbf{E}}\right \rangle_0}
\end{equation}
a weighted sum in the absence of the field. The average is over a joint equilibrium distribution that can be obtained by reweighting its nonequilibrium counterpart 
$p_E(\mathbf{J},\mathbf{Q})=\langle \delta(J_{\textrm{i}}-J_{\textrm{i}}[\ve{X}(\tN)],Q_{\textrm{i}}-Q_{\textrm{i}}[\ve{X}(\tN)],J_{\textrm{w}}-J_{\textrm{w}}[\ve{X}(\tN)],Q_{\textrm{w}}-Q_{\textrm{w}}[\ve{X}(\tN)]) \rangle_E$ according to the equation
\begin{equation}
\label{Eq:ReweightP_Explicit}
\frac{\ln p_0(\mathbf{J},\mathbf{Q})}{\tN} = \frac{\ln
p_E(\mathbf{J},\mathbf{Q})}{\tN}- \beta(\mathbf{J}+\mathbf{Q}) \cdot
\frac{\mathbf{E}}{2} +\frac{\beta}{4} \boldsymbol{N\lambda} \cdot
\mathbf{E}^{(2)}
\end{equation}
as before.  Note that in Eq.~\ref{Eq:Current_Explicit}, the dynamical quantities of the water are coupled with the ions through the relative action. Even for observables that depend only on a subset of degrees of freedom, like those of the ions, the exponential bias correlates them with the entire set of degrees of freedom in the system.

A practical difficulty arises in applying Eq.~\ref{Eq:ReweightP_Explicit} straightforwardly. As $\mathbf{J}$ and $\mathbf{Q}$ are both extensive variables in particle number and observation time, when both are large, $p_E(\mathbf{J},\mathbf{Q})$ becomes exponentially peaked about its most typical values, and the reweighting factors become large enough that it is difficult to represent them numerically. This makes performing the reweighting cumbersome. A solution is found for observables that depend only on the ion degrees of freedom, by treating the contribution to the reweighting factor from the water approximately. An accurate approximation can be found by  first writing the joint distribution using Bayes theorem, 
\begin{eqnarray}
p_E(\mathbf{J},\mathbf{Q})&=&p_E(J_{\textrm{i}},Q_{\textrm{i}} | J_{\textrm{w}},Q_{\textrm{w}})p_E(J_{\textrm{w}},Q_{\textrm{w}}) \\
&=& p_E(J_{\textrm{i}},Q_{\textrm{i}} | J_{\textrm{w}},Q_{\textrm{w}}) \exp[ - \Nw\tN I_E(\bar{j}_{\textrm{w}},\bar{q\vphantom{j}}_{\textrm{w}})] \nonumber
\end{eqnarray}
where the first term on the right hand side is a conditional probability, and in the second line $I_E(\bar{j}_{\textrm{w}},\bar{q\vphantom{j}}_{\textrm{w}})$ is the rate function for intensive variables $\bar{j}_{\mathrm{w}} = J_{\textrm{w}}/ N_{\mathrm{w}}$ and $\bar{q\vphantom{j}}_{\textrm{w}}=Q_{\textrm{w}}/ N_{\mathrm{w}}$. The form of the marginal distribution of the water variables is known as a large deviation form, and holds in the limit of large number of particles and observation time.  
Under the assumption that the joint distribution $I_E(\bar{j}_{\textrm{w}},\bar{q\vphantom{j}}_{\textrm{w}})$ is peaked in $(\bar{j}_{\textrm{w}},\bar{q\vphantom{j}}_{\textrm{w}})$, we  use a saddle point approximation to evaluate their contribution to the reweighting factors in Eq.~\ref{Eq:Current_Explicit}. For  the marginal distribution of the ion variables $p_0(J_{\mathrm{i}},Q_{\mathrm{i}})$ this leads to an approximate reweighting,
\begin{equation}
\begin{aligned}
\label{Eq:ReweightP_approx}
\ln p_0(J_{\mathrm{i}},Q_{\mathrm{i}}) &\approx \ln p_E[J_{\mathrm{i}},Q_{\mathrm{i}}|j_{\textrm{w}}^*(E_{\textrm{w}}),q_{\textrm{w}}^*(E_{\textrm{w}})] \\
&- \beta \tN (J_{\mathrm{i}}+Q_{\mathrm{i}}) \frac{{E}_{\mathrm{i}}}{2} -\mathcal{N}(E_{\mathrm{i}},E_{\mathrm{w}})
\end{aligned}
\end{equation}
where $[j_{\textrm{w}}^*(E),q_{\textrm{w}}^*(E)]$ denotes the maximizer
\begin{equation}
[ j_{\textrm{w}}^*(E),q_{\textrm{w}}^*(E) ] =\underset{(\bar{j}_{\textrm{w}},\bar{q\vphantom{j}}_{\textrm{w}})}{\textrm{argmin}} \left \{\,I_E(\bar{j}_{\textrm{w}},\bar{q\vphantom{j}}_{\textrm{w}})+\beta (\bar{j}_{\textrm{w}}+\bar{q\vphantom{j}}_{\textrm{w}})E/2 \right \}
\end{equation}
and the normalization constant $\mathcal{N}(E_{\mathrm{i}},E_{\mathrm{w}}) $ becomes
\begin{equation}
\mathcal{N}= \Nw \tN I_{E_{\textrm{w}}}(j^*_{\textrm{w}},q^*_{\textrm{w}})
+ \frac{\beta}{2} \Nw \tN (j^*_{\textrm{w}}+q^*_{\textrm{w}})E_{\textrm{w}} - \frac{\beta}{4}
\tN \boldsymbol{N \lambda} \cdot \mathbf{E}^{(2)}
\end{equation}
which depends only on the applied external fields, not on any fluctuating variables. 

We use this approximate expression for the reweighting procedure in the results
presented below. As the extensive dynamical variables related to water molecules
are usually larger in magnitude compared to the ions by an order of magnitude,
due to $\Nw \gg N_{\mathrm{i}}$, this expression greatly reduces numerical
issues in dealing with the exponential of very large numbers. Note that when the
rate function $I_E(\bar{j}_{\textrm{w}},\bar{q\vphantom{j}}_{\textrm{w}})$ is quadratic, or the ions and
water are uncorrelated, this approximation is exact. Outside of those regimes, we still find it admits a faithful approximation to the exact reweighting relation in Eq.~\ref{Eq:ReweightP_Explicit}, especially when using additional fields to reconstruct $\ln p_0(J_{\mathrm{i}},Q_{\mathrm{i}})$ far into the tails of the distribution.  

To compute the molar ionic conductivity, we can straightforwardly adapt Eq.~\ref{Eq:Reweightlam0}. We first consider the physical condition where the same field  $E_{\textrm{i}}=E_{\textrm{w}}=E$ is applied on the whole system. The ionic conductivity defined as $\lambda(E)=(1/N_{\textrm{i}})d\langle J_{\textrm{i}}\rangle_E/dE$ can be computed as
\begin{eqnarray}
\label{Eq:Conductivity_Fieldonboth}
\lambda(E)=&&\lim_{t_{\textrm{N}}\rightarrow\infty} \frac{\beta t_{\textrm{N}}}{2 N_{\textrm{i}}}\left\langle(\delta J_{\textrm{i}}^2+\delta J_{\textrm{i}}\delta Q_{\textrm{i}}+\delta J_{\textrm{i}}\delta J_{\textrm{w}} \right . \nonumber\\
&&\left .+\delta J_{\textrm{i}}\delta Q_{\textrm{w}})e^{\beta \Delta U_{\ve{E}}[\ve{X}(\tN)]}\right\rangle_0
\end{eqnarray}
which includes both the self-correlations among the ions, and the cross-correlations between ions and water molecules. This becomes evident upon a Taylor expansion on Eq.~\ref{Eq:Conductivity_Fieldonboth} to second order in the applied field, where the long time limit is implied but suppressed for brevity, 
\begin{equation}
\label{Eq:Taylor_Fieldonboth}
\begin{aligned}
&\lambda(E)=\frac{\beta\tN}{2N_{\textrm{i}}}\left(\left\langle J_{\textrm{i}}^2 \right\rangle_0 + \left\langle J_{\textrm{i}} J_{\textrm{w}} \right\rangle_0 \right) \\
&+\frac{3\beta^3\tN^3}{8N_{\textrm{i}}}\left( \frac{1}{6}\left\langle J_{\textrm{i}}^4 \right\rangle_0 - \frac{1}{2} \left\langle J_{\textrm{i}}^2 \right\rangle_0^2 + \frac{1}{2} \left\langle \delta(J_{\textrm{i}}^2)\delta(Q_{\textrm{i}}^2) \right\rangle_0 \right. \\
&+\frac{1}{2}\left\langle J_{\textrm{i}}^3J_{\textrm{w}} \right\rangle_0 - \frac{3}{2} \left\langle J_{\textrm{i}}J_{\textrm{w}} \right\rangle_0\left\langle J_{\textrm{i}}^2 \right\rangle_0 + \left\langle \delta(J_{\textrm{i}}^2)\delta(Q_{\textrm{i}}Q_{\textrm{w}}) \right\rangle_0 \\
&+\frac{1}{2}\left\langle\delta(J_{\textrm{i}}J_{\textrm{w}}) \delta(Q_{\textrm{i}}^2) \right\rangle_0 + \frac{1}{2}\left\langle \delta(J_{\textrm{i}}^2)\delta(J_{\textrm{w}}^2)) \right\rangle_0 - \left\langle J_{\textrm{i}}J_{\textrm{w}} \right\rangle_0^2 \\
&+\frac{1}{2}\left\langle \delta(J_{\textrm{i}}^2)\delta(Q_{\textrm{w}}^2) \right\rangle_0 + \left\langle \delta(J_{\textrm{i}}J_{\textrm{w}})\delta(Q_{\textrm{i}}Q_{\textrm{w}}) \right\rangle_0 + \frac{1}{6}\left\langle J_{\textrm{i}}J_{\textrm{w}}^3\right\rangle_0\\
&\left. -\frac{1}{2}\left\langle J_{\textrm{i}}J_{\textrm{w}} \right\rangle_0\left\langle J_{\textrm{i}}^2 \right\rangle_0 + \frac{1}{2}\left\langle \delta(J_{\textrm{i}}J_{\textrm{w}}) \delta(Q_{\textrm{w}}^2)\right\rangle_0 \right)E^2+O(E^4)
\end{aligned}
\end{equation}
where time-reversal and spatial symmetry are invoked to eliminate terms of zero value.

Alternatively, one can construct an artificial perturbation where $E_{\mathrm{i}}=E$ and $E_{\mathrm{w}}=0$, where the external field is only applied to the ions. While this condition is not physical, its utility is illustrated in the expression for molar ionic conductivity,
\begin{equation}
\label{Eq:Conductivity_Fieldonion}
\tilde{\lambda}(E_i)=\lim_{t_{\textrm{N}}\rightarrow\infty} \frac{\beta t_{\textrm{N}}}{2N_{\textrm{i}}}\left\langle(\delta J_{\textrm{i}}^2+\delta J_{\textrm{i}}\delta Q_{\textrm{i}})e^{\beta \Delta U_{\ve{E}_i}[\ve{X}(\tN)]}\right\rangle_0
\end{equation}
where compared to Eq.~\ref{Eq:Conductivity_Fieldonboth}, all the cross-correlations between the ions and the water disappear in the Taylor expansion, where again the long time limit is implied
\begin{equation}
\label{Eq:Taylor_Fieldonion}
\begin{aligned}
\tilde{\lambda}(E_i)&=\frac{\beta\tN}{2N_{\textrm{i}}}\left\langle J_{\textrm{i}}^2 \right\rangle_0 +\frac{3\beta^3\tN^3}{8N_{\textrm{i}}}\left( \frac{1}{6}\left\langle J_{\textrm{i}}^4 \right\rangle_0 - \frac{1}{2} \left\langle J_{\textrm{i}}^2 \right\rangle_0^2 \right.\\
&+ \left. \frac{1}{2} \left\langle \delta(J_{\textrm{i}}^2)\delta(Q_{\textrm{i}}^2) \right\rangle_0\right)E_i^2+O(E_i^4)
\end{aligned}
\end{equation}
Thus by comparing the conductivity under the two scenarios, we can obtain an estimate of the contribution from ion-water correlations, which is not readily extracted from direct simulations and integrations of correlations functions.

\begin{figure}[t]
\begin{center}
\includegraphics[width=8.5cm]{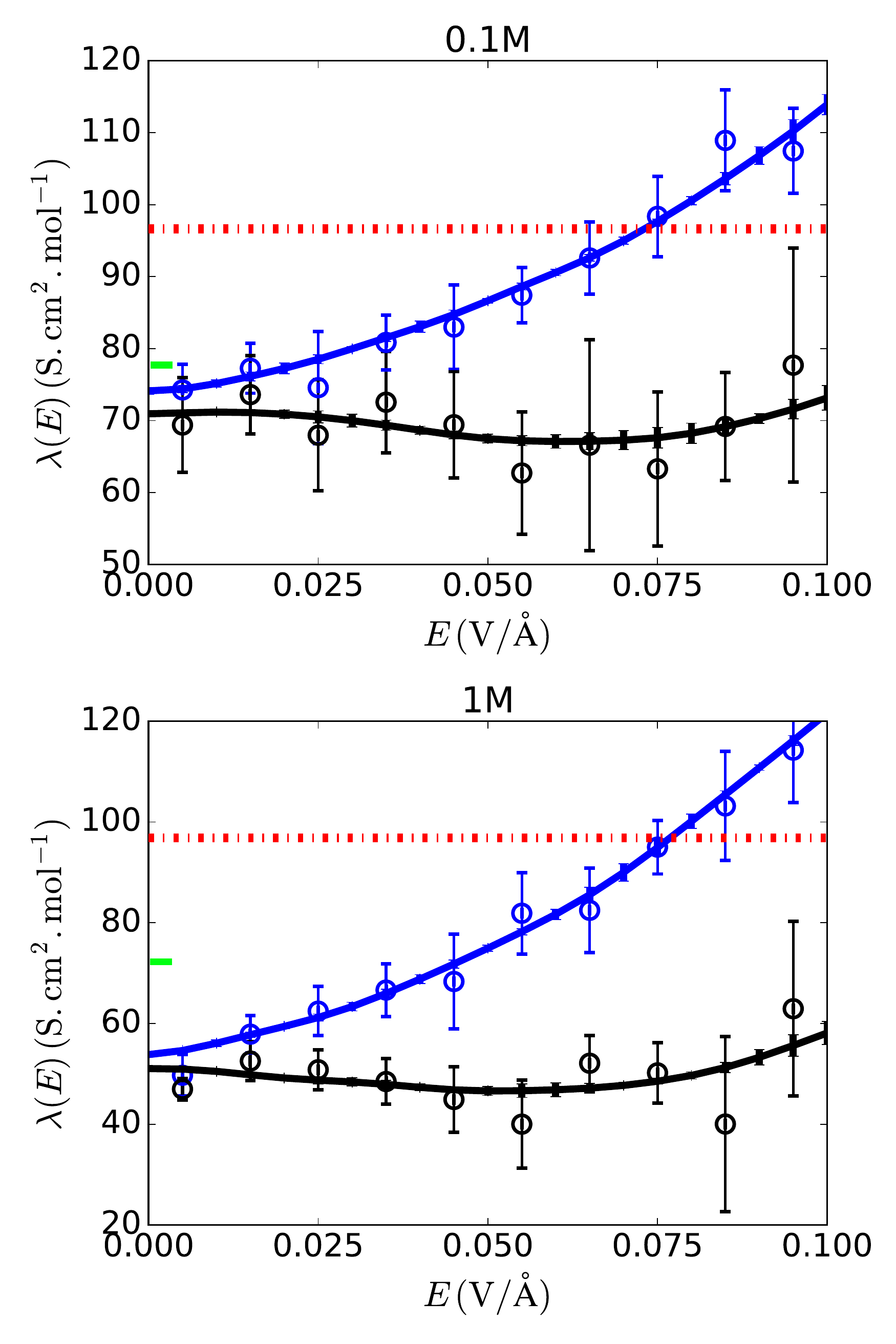}
\caption{Field-dependent conductivities of the explicit model where an external field is applied on both the ions and water (black), and only applied on the ions (blue). The lines are computed from the reweighting method, while the symbols are computed from finite differences of $\left\langle J_{\textrm{i}}\right\rangle_E$ versus $E$, with one standard deviation as errorbars. The top are computed at 0.1 M and bottom 1.0 M. Green tick marks on the left axis denote the MSA prediction for $E=0$.}
\label{fig:Explicit}
\end{center} 
\end{figure}
\subsection{Ion-water correlations suppress nonlinear response}

Shown in Fig.~\ref{fig:Explicit} are the field-dependent conductivities of the explicit solvent system at two concentrations, 0.1M and 1M. The reweighted results for both are computed from Eq.~\ref{Eq:ReweightP_approx} using $\tN=10\textrm{fs}$, which is long enough to justify the saddle point approximation and converge the mean reweighted current. Specifically, we first constructed $p_0(J_{\textrm{i}}, Q_{\textrm{i}})$ using a series of simulations at finite fields at the locations of the direct estimate in Fig.~\ref{fig:Explicit}, and the generalized version of WHAM to stitch joint histograms of $J_{\textrm{i}}$ and $Q_{\textrm{i}}$ together. Additionally, we have computed the conductivity from a numerical derivative of the average current directly from a set of simulations at fixed field. While the two estimates are in good quantitative agreement, the statistical errors are much smaller for the reweighted results, as data across the whole fields are supplemented in each estimate. 

We find the conductivity is only weakly dependent on the field and the curve is
well fitted by a polynomial with a positive fourth order and a negative second
order term. The curves for the two different concentrations, 0.1M and 1M are
remarkably similar, though the latter exhibits a consistently lower
conductivity.  The conductivity at zero field for both concentrations is
suppressed relative to its value at infinite dilution, or compared to its
implicit solvent value.  For 0.1M, the ionic molar conductivity at zero field
agrees well with the result from the mean spherical approximation (MSA, see
App.~\ref{Sec:MSA}),\cite{Turq2005} which is equal to 78 S cm$^2/$mol using ionic radii
$r_{\textrm{c}}=3.05$\AA~and $r_{\textrm{a}}=3.62$\AA~). This agreement implies that at 0.1M, the dynamical response of the ions is well described by Gaussian fluctuations in the density and velocity fields.\cite{chandler1993gaussian,speck2013gaussian} 
More details are available in App.~\ref{Sec:MSA}. MSA however overpredicts the conductivity at 1.0 M with a value of 72 S cm$^2/$mol, reflecting the underestimation of ionic correlation effects.
At unphysically high fields, there is a significant nonlinear response due to the dielectric saturation of the solvent, see App.~\ref{Sec:HighE}.

From the ensemble reweighting theory presented in the previous section, we have
a means of decoupling the contributions to the field-dependent conductivity from
the solvent and those from the ions. Specifically, we can use a generalized
ensemble where only a field is applied to the ions, not the water, to deduce
which correlations suppress the field-dependence that result directly from the water.    
When the field is only applied on the ions for both concentrations,
the conductivity grows quadratically with a large positive second order term.
This is shown in Fig.~\ref{fig:Explicit}. The drastic difference between these
two sets of results can be unravelled by a comparison between the generalized
fluctuation-dissipation relationships in Eq.~\ref{Eq:Taylor_Fieldonboth} and
Eq.~\ref{Eq:Taylor_Fieldonion}, where all cross-correlations between the ions
and water are absent from the latter expression. More specifically, the lower
conductivity at zero field when the field is applied on both the ions and water
is a direct result of negative correlations in $\left\langle
J_{\textrm{i}}J_{\textrm{w}}\right\rangle_0$ at equilibrium. The negative second
order coefficient results from the negative fourth order correlations between
ions and water, among which the dominant term is $\left\langle \delta(J_{\textrm{i}}^2) \delta(Q_{\textrm{w}}^2)\right\rangle_0$ 
due to the larger number of water molecules and subsequently larger fluctuations in $Q_{\textrm{w}}$.

\begin{figure}
\begin{center}
\includegraphics[width=8.5cm]{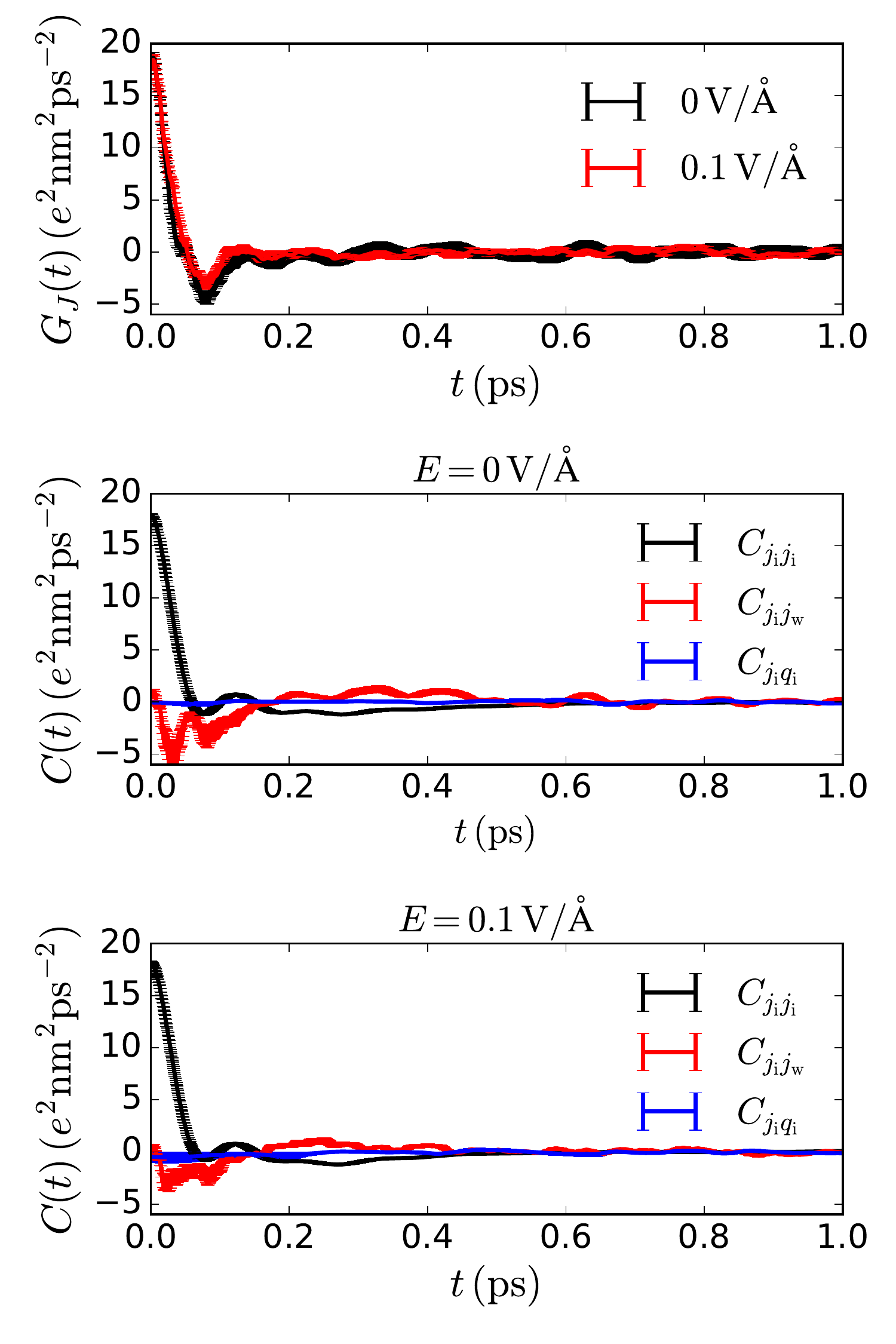}
\caption{Time correlation functions for field-dependent conductivities of 0.1 M explicit solvent solution. (Top) time correlation functions $G_J(t)=C_{ j_{\textrm{i}} j_{\textrm{i}}}(t)+C_{ j_{\textrm{i}} j_{\textrm{w}}}(t)$ for $E=0$ and
	$E=0.1 \, $V$/\mathrm{\AA}$. (Middle) Contributions to the conductivity
from current-current and current-frenesy for $E=0 \, $V$/\mathrm{\AA}$. (Bottom)
Contributions to the conductivity from current-current and current-frenesy for $E=0.1 \, $V$/\mathrm{\AA}$. Throughout errorbars are one standard deviation from the mean.
}
\label{fig:Corr_explicit}
\end{center} 
\end{figure}
To gain more physical insight into the relevant correlations, we rewrite the equation for the conductivity using time integrated correlation functions
\begin{equation}
\begin{aligned}
\lambda(E)&=\lim_{\tN\rightarrow\infty}\frac{\beta\tN}{2N_{\textrm{i}}}\left\langle \delta J_{\textrm{i}}^2+\delta J_{\textrm{i}}\delta J_{\textrm{w}}+\delta J_{\textrm{i}}\delta Q_{\textrm{i}}+\delta J_{\textrm{i}}\delta Q_{\textrm{w}} \right\rangle_E\\
&=\frac{\beta}{N_{\textrm{i}}}\int_0^{\infty}dt \,\,\left[ G_J(t)+G_Q(t) \right]\\
\end{aligned}
\end{equation}
where $G_J(t)=C_{ j_{\textrm{i}} j_{\textrm{i}}}(t)+C_{ j_{\textrm{i}} j_{\textrm{w}}}(t)$, with
\begin{equation}
\begin{aligned}
C_{ j_{\textrm{i}} j_{\textrm{i}}}(t)&=\left\langle\delta j_{\textrm{i}}(0)\delta j_{\textrm{i}}(t)\right\rangle_E\\
C_{ j_{\textrm{i}} j_{\textrm{w}}}(t)&=\frac{1}{2}\left\langle\delta j_{\textrm{i}}(0)\delta j_{\textrm{w}}(t)+\delta j_{\textrm{i}}(0)\delta j_{\textrm{w}}(-t)\right\rangle_E
\end{aligned}
\end{equation}
and $G_Q(t)=C_{ j_{\textrm{i}} q_{\textrm{i}}}(t)+C_{ j_{\textrm{i}} q_{\textrm{w}}}(t)$, with  
\begin{equation}
\begin{aligned}
C_{ j_{\textrm{i}} q_{\textrm{i}}}(t)&=\frac{1}{2}\left\langle\delta j_{\textrm{i}}(0)\delta q_{\textrm{i}}(t)+\delta j_{\textrm{i}}(0)\delta q_{\textrm{i}}(-t)\right\rangle_E \\
C_{ j_{\textrm{i}} q_{\textrm{w}}}(t)&=\frac{1}{2}\left\langle\delta j_{\textrm{i}}(0)\delta q_{\textrm{w}}(t)+\delta j_{\textrm{i}}(0)\delta q_{\textrm{w}}(-t)\right\rangle_E
\end{aligned}
\end{equation}
The conductivity away from equilibrium is a sum of the integrated
current-current correlation function, denoted by $G_J$, and integrated
current-frenesy correlation function, denoted by $G_Q$. At zero field,  $G_Q$ is
zero due to time reversal symmetry, leaving $G_J$ as the only contribution to
the zero-field conductivity. At finite fields when $G_Q$ should contribute to
the conductivity as well, for the concentrations studied, we find that the
contribution from $C_{ j_{\textrm{i}} q_{\textrm{i}}}(t)$ is negligible compared to
the other three terms. This is due to the screening effect of the explicit water
molecules that results in the ions being largely dissociated, so that the
effect of ion-ion interactions arises mainly from the relaxation of the ionic cloud 
rather than from ion pairing and is thus weaker than short-range ion-water interactions. 
The $C_{ j_{\textrm{i}} q_{\textrm{w}}}(t)$ term exhibits large
statistical fluctuations due to the larger number and stronger intramolecular
forces of the water molecules, and is thus much more difficult to converge by
brute force calculations. We have to infer its contribution from the measured
conductivities using the generalized fluctuation-dissipation relationships.

Figure~\ref{fig:Corr_explicit} shows the current-current contribution to the
response function for the 0.1M explicit system at equilibrium $E=0$ and at a
finite field $E=0.1  \mathrm{V/\AA}$. Also in Fig.~\ref{fig:Corr_explicit}, the
total correlation function for $E=0$ and $E=0.1 \mathrm{V/\AA}$ are decomposed
into their various pieces.  At zero field, both the ion-ion current
self-correlation and the ion-water current cross-correlation exhibit noticeable
recoil effects evident in transient negative correlations. The former (ion-ion),
which spreads over longer timescales, integrates to a positive contribution, and
is the only term responsible for the zero-field conductivity when the field is
only applied to the ions. The latter (ion-water) integrates to a much smaller
negative contribution, which accounts for the difference between the zero-field
conductivities between the two sets in Fig.~\ref{fig:Explicit}. At a higher
field $E=0.1~\mathrm{V/\AA}$, the recoil effect in both correlations is
reduced, resulting in a larger integrated value of both current-current
correlation functions, and a more positive $G_J(t)$. 

We can infer the large contribution from $C_{ j_{\textrm{i}} q_{\textrm{w}}}(t)$ at finite field by contrasting the behavior of $\lambda(E)$ and $\tilde{\lambda}(E)$. 
In the case where the field is applied only on the ions, the correlation function
$C_{ j_{\textrm{i}} j_{\textrm{i}}}(t)$ integrates to a similarly large contribution as in the case when the field is applied to both ions and water, and one that is larger than its zero field value. As  $C_{ j_{\textrm{i}} j_{\textrm{w}}}(t)$ and $C_{ j_{\textrm{i}} q_{\textrm{i}}}(t)$  are persistently small at $E=0.1\mathrm{V/\AA}$, the large difference difference between $\lambda(E)$ and $\tilde{\lambda}(E)$ must result from significant negative contributions in the 
current-frenesy correlation function $C_{ j_{\textrm{i}} q_{\textrm{w}}}(t)$ between the ions and the water. The physical origin of this negative correlation is the
relaxation effects that result from ion displacements that transiently distort
the local dielectric environment and generate a restoring force on the water
molecules from the compensating ionic cloud left behind.
This term dominates the total correlation function and compensates the increase
in the current-current correlation functions, yielding a weak dependence on
field of the conductivity. Thus while the current-frenesy correlation function is
difficult to compute explicitly, from the above analysis we are able to infer its effect using the decomposition of time correlation functions.

\begin{figure}
\begin{center}
\includegraphics[width=8.5cm]{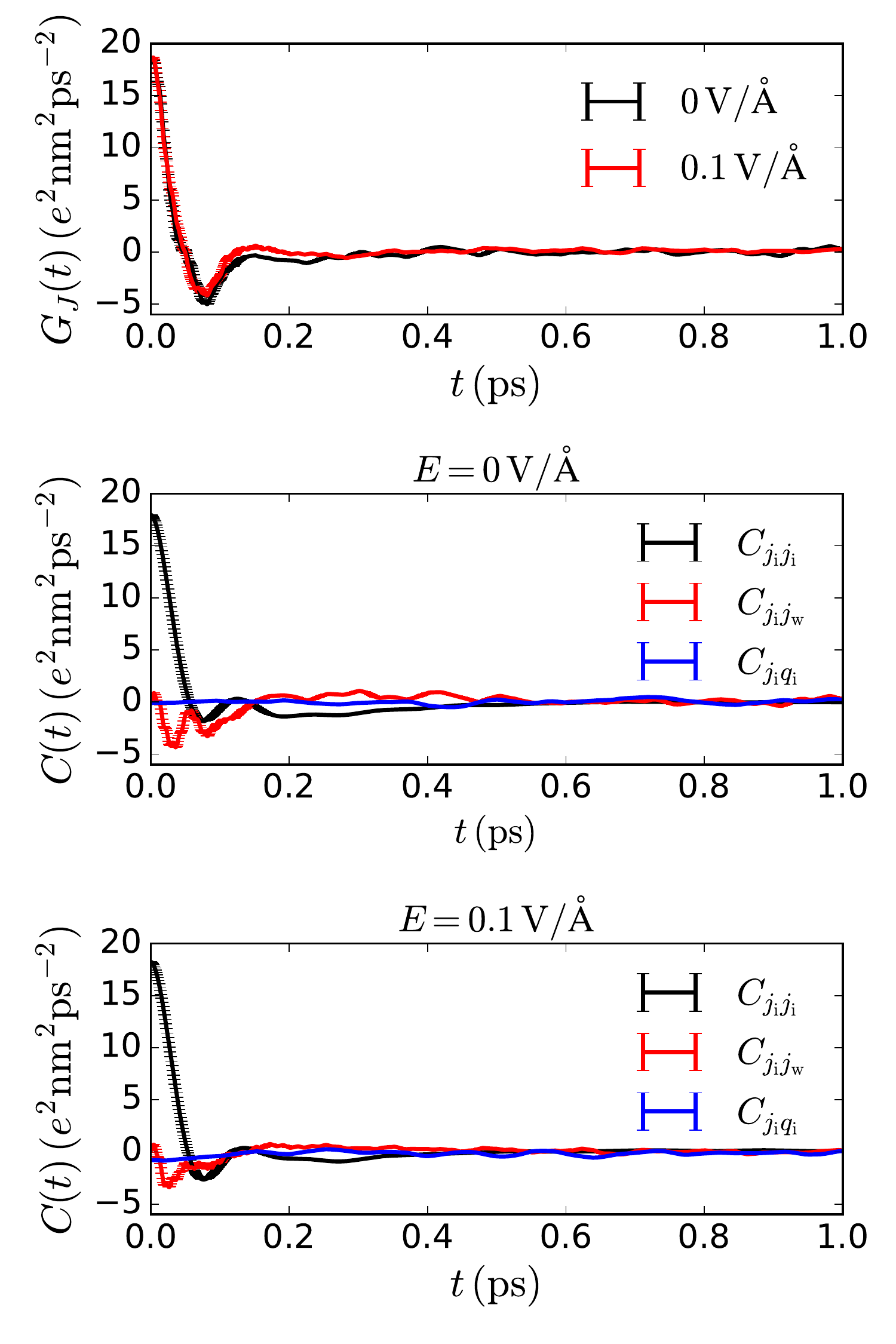}
\caption{Time correlation functions for field-dependent conductivities of 1.0 M explicit solvent solution. (Top) time correlation functions $G_J(t)=C_{ j_{\textrm{i}} j_{\textrm{i}}}(t)+C_{ j_{\textrm{i}} j_{\textrm{w}}}(t)$ for $E=0$ and
	$E=0.1 \, $V$/\mathrm{\AA}$. (Middle) Contributions to the conductivity
from current-current and current-frenesy for $E=0 \, $V$/\mathrm{\AA}$. (Bottom)
Contributions to the conductivity from current-current and current-frenesy for $E=0.1 \, $V$/\mathrm{\AA}$. Throughout errorbars are one standard deviation from the mean.
}
\label{fig:Corr_explicit2}
\end{center} 
\end{figure}
Figure ~\ref{fig:Corr_explicit2} shows the correlation functions for the 1.0~M
system. Qualitatively, they are very similar to those for the 0.1~M solution.
Both systems exhibit noticeable recoil effects in the current-current function
due to ion-ion terms. In the 1.0~M solution, the transient negative correlation
is larger than in the 0.1~M solution. This is also true for the  current-current
function due to ion-water terms. The combination of these negative correlations
results in the reduced conductivity at $E=0~\mathrm{V/\AA}$. As for the small concentration,
the currrent-frenesy correlations from ion-ion terms are negligible even at 0.1~V/$\mathrm{\AA}$, and the suppressed field-dependence results from the
currrent-frenesy correlations from ion-water terms. 

\section{Comparison between implicit and explicit solvents}
The ability to efficiently compute field-dependent conductivities generally
enables a comparison between different solvent representations. 
Comparing the implicit and explicit models, we find that at the same
concentration of 0.1 M the conductivity in the implicit model is higher by
nearly  30 \%. The implicit model has been parameterized to yield the correct
limiting conductivity in the infinite dilution regime, so the enhanced
conductivity in the implicit model at zero applied field results from an
inaccurate description of ion-ion correlations. It is well known that
inter-ionic correlations are not accurately described by screening by a lone
dielectric constant, as explicit solvent models induce potentials of mean force
between ions that exhibit significant structure.\cite{pettitt1986alkali} 
Additionally, in our study only the explicit model considers hydrodynamic effects that result from momentum transfer and counter flows around the ion. 
When we neglect hydrodynamic effects for the implicit solvent, we find consistent agreement between Onsager's theory and MSA for the strong electrolyte.
Indeed, in the dilute limit MSA reduces to Wilson's theory. However, neither linear theory is capable of describing the weak electrolyte, as both MSA and Onsager's theory erroneously predict a negative conductivity.  MSA is capable of qualitatively predicting the reduced conductivity with concentration for the strong electrolyte, but is not quantitatively accurate. 
 
 With application of an electric
field, we find that both the implicit and explicit models have a nearly constant
conductivity for this strong electrolyte system, but for different reasons. The
implicit model saturates its conductivity to the Nernst-Einstein value at zero
field. Increasing the field has little effect on the already weak ion
correlations. The nearly constant conductivity in the explicit model, however,
results from a balancing of decreased inter-ionic correlations that act to
increase the conductivity with increased ion-water correlations that suppress
it. In other models, there is no reason that these need to be so carefully balanced,
suggesting that care should be taken in interpreting results of implicit models
under large applied fields.

\section{Conclusion}
We have shown how to relate dynamical fluctuations of molecular quantities
within nonequilibrium steady states to observable macroscopic response for
detailed molecular models of electrolyte solutions. Such calculations are made
possible formally by advances in the theory of nonequilibrium steady
states\cite{gao2018nonlinear} and numerically by employing an ensemble
reweighting relation that enables an importance sampling of a probability
distribution of time integrated functions.\cite{lesnicki2020field} Considering
the field-dependence of the ionic conductivity, we have shown how nonlinear
relationships between an ionic current and applied electric field can emerge in
weak electrolytes as ion correlations are reduced, and how they are mitigated in
strong electrolytes due to persistent solvent friction. Whereas for strong
electrolytes at low concentrations, existing approximate
theories\cite{wilson1936theory,Turq2005} can accurately predict the conductivity
and its field-dependence, for weak electrolytes or elevated concentrations these
theories break down. 

In the explicit solvent case, the present theoretical and
numerical techniques also provide new opportunities to investigate coupled charge and mass 
transport in electrolytes\cite{sedlmeier_charge/mass_2014} or the origin of the 
frequency-dependent solvent friction on the ions and its consequences on the
conductivity of electrolyte solutions.\cite{chandra_frequency_2000,dufreche_ionic_2002,balos_macroscopic_2020}
The present approach with explicit solvent can also be used to test theories beyond the hypotheses underlying that of Onsager and Wilson, such as based on lattice models,\cite{eyink1996hydrodynamics} fluctuating hydrodynamics\cite{peraud2017fluctuation,donev2019fluctuating} or stochastic density functional theory.\cite{demery2016conductivity}
While we have considered bulk solutions of monovalent electrolytes, our
approach is straightforwardly applied for multivalent ions, where nonlinear
responses due to field-induced ion pair dissociation should be more prominent,
and can be extended to instances of transport in confinement,\cite{pean2015confinement,
simonnin2018mineral,palmer2020correlation, jin2020coupling,robin2021principles} in which case
generalizations of Eq.~\ref{Eq:GrE} for nonlinear response may provide insight
into recent experimental observations.\cite{esfandiar2017size} Further, the efficient evaluation of the field-dependent conductivity could be straightforwardly applied to complex fluids like ion transport in polymer electrolytes, which may provide design principles for energy storage devices.\cite{wettstein2021polymer}

\section*{Data availability} 
The data that support the findings of this study are openly available in Zenodo at https://doi.org/10.5281/zenodo.4660193\cite{data}

\section*{Acknowledgements}
D. L. and B. R. acknowledge financial support from the French Agence Nationale de la Recherche (ANR) under
Grant No. ANR-17-CE09-0046-02 (NEPTUNE) and from the H2020-FETOPEN project NANOPHLOW (Grant No. 766972).
B. R. acknowledges financial support from the European Research Council (ERC) under the European Union's Horizon 2020 research and innovation programme (grant agreement No. 863473).
D. T. L. and C. Y. G. was supported by the LDRD program at LBNL under U. S. Department of Energy Office of Science, Office of Basic Energy Sciences under Contract No. DE-AC02-05CH11231.

\appendix

\section{Onsager-Wilson theory}
\label{Sec:Wilson}

In the main text, we compare our results with Onsager-Wilson analytical theory for the
field-dependent conductivity. The theory considers both a relaxation effect, derived from the
perturbation of the ionic atmosphere in the presence of the field, and
hydrodynamic effect, calculated by solving the Navier-Stokes equation.
For a symmetric electrolyte consisting of point ions with charges $\pm|q|e$
and concentration $\bar{\rho}$
in a solvent with relative permittivity $\varepsilon$ and viscosity $\eta$,
the conductivity is given by~\cite{harned1959physical}:
\begin{equation}
\label{Eq:Wilson}
\sigma(E) =  \sigma_0 -\Delta\sigma_{\mathrm{rel}}(\beta |q|eE/ \kappa)
-\Delta\sigma_{\mathrm{hyd}}(\beta |q|eE/ \kappa)
\end{equation}
with $\sigma_0$ the Nernst-Einstein conductivity. The second and third terms account for the relaxation of the ionic cloud and for hydrodynamic interactions between ions, respectively.
These depend naturally on the reduced field $x=\beta |q|eE/ \kappa$, with
$\kappa$ the inverse of the Debye screening length.
Specifically, the two contributions are
\begin{equation}
\Delta\sigma_{\mathrm{rel}}(x) = \sigma_0 \frac{\kappa \ellB}{2}  h(x)
\end{equation}
and
\begin{equation}
\Delta\sigma_{\mathrm{hyd}}(x) = \frac{ 2 \bar{\rho}|q|e }{6\sqrt{2}\pi  \eta}f(x)
\end{equation}
with two smooth functions
\begin{eqnarray}
h(x) &=& -\frac{1}{2x^3} \left[ -x\sqrt{1+x^2} + \tan^{-1}\frac{x}{\sqrt{1+x^2}} \right . \nonumber \\
&&\left . + \sqrt{2}x - \tan^{-1}\sqrt{2}x\right] \, .
\end{eqnarray}
and 
\begin{eqnarray}
&f(x) = 1 + \frac{3}{4\sqrt{2}x^3} \left[  2x^2\sinh^{-1}x -x\sqrt{1+x^2} + \sqrt{2}x \right . \nonumber \\
&\left.  - (1+2x^2)\tan^{-1}\sqrt{2}x + (1+2x^2)\tan^{-1}\frac{x}{\sqrt{1+x^2}} \right]
\end{eqnarray}
representing the contributions form the ion relaxation and hydrodynamic effects, respectively. Full equations for the field-dependent pair distribution function can be found in Ref.~\onlinecite{eu2010onsager}. 

\section{Mean spherical approximation}
\label{Sec:MSA}

We report here the expression for the conductivity at equilibrium ($E=0$) 
within the mean spherical approximation (MSA)\cite{lebowitz1966mean}. MSA
is a refinement over the Onsager-Fuoss theory, considering excluded volume
effects on both electrostatic and hydrodynamic interactions. We consider an electrolyte consisting of
ions with diameters $\sigma_i$, charges $z_i=q_ie$ and concentrations $\bar{\rho}_i$ (in $m^{-3}$)
in a solvent with relative permittivity $\varepsilon$ and viscosity $\eta$:~\cite{Turq2005}
\begin{equation}
\sigma_{\mathrm{MSA}} = \sum_{ij} \bar{\rho}_i q_iq_j \frac{L_{ij}}{T}
\end{equation}
where $L_{ij}$ are the Onsager coefficients defined by
\begin{equation}\label{Eq:MSA}
\frac{L_{ij}}{T} = \left(\frac{D_i}{\kB T}\delta_{ij} + {\Omega_{ij}} \right)
\left( 1 +\frac{\delta F_j^{rel}}{F_j^{ext}} \right)
\end{equation}
with $\Omega_{ij}$ related to the hydrodynamic interactions between the
particles and $\frac{\delta F_j^{rel}}{F_j^{ext}}$ the relaxation term.
The former is obtained from the equilibrium distribution $h_{ij}^0(r)$ function 
\begin{equation}
\Omega_{ij} = \frac{2}{3\eta} \bar{\rho}_j \int_0^\infty rh_{ij}^0(r)dr
\; .
\end{equation}
Expanding $h_{ij}^0(r)$ within the MSA into several contributions 
results in the following expression:
\begin{equation}
\Omega_{ij} = \Omega^{HS}_{ij} + \Omega^{c1}_{ij} + \Omega^{c2}_{ij}
\end{equation} 
where $\Omega^{HS}_{ij}$ is the hard sphere electrophoretic term,
$\Omega^{c1}_{ij}$ the MSA electrostatic term,
$\Omega^{c2}_{ij}$ the second-order electrostatic term.
The hard-sphere electrophoretic term is given by
\begin{equation}
\Omega^{HS}_{ij} = - \frac{(\sigma_j + \sigma_j)^2}{12 \eta} \bar{\rho}_j 
\frac{1-\tilde{X}_3/5 +\tilde{X}_3^2/10 }{1+2\tilde{X}_3}
\end{equation}
with $\tilde{X}_3 = \frac{\pi}{6} \sum \bar{\rho}_i \frac{3 X_1 X_2/ X_0 + X_3}{4 X_0}$ 
where $X_n = \frac{\pi}{6}  \sum \bar{\rho}_i\sigma^n_i$.
Introducing the screening parameter $\Gamma$,
\begin{equation}
\Gamma = \left[ \pi \ellB \sum_i \bar{\rho}_i
\left(\frac{q_i}{1+\Gamma\sigma_i}\right)^2\right]^{1/2}
\end{equation}
the MSA electrostatic term is	
\begin{equation}
\Omega^{c1}_{ij} = - \frac{1}{3\eta} \frac{q_iq_j\ellB
\bar{\rho}_j}{(1+\Gamma\sigma_i)(1+\Gamma\sigma_j)\left( \Gamma + \sum_k
\bar{\rho}_k \frac{\pi \ellB q_k^2\sigma_k}{(1+\Gamma\sigma_k)^2}\right)}
\end{equation}
and the second-order electrostatic term is
\begin{equation}
\Omega^{c2}_{ij} =
\frac{\ellB^2\bar{\rho}_j^2q_i^2q_j^2}{3\eta(1+\Gamma\sigma_i)(1+\Gamma\sigma_j)}
e^{2\kappa \sigma_{ij}}\int_{2\kappa \sigma_{ij}}^{\infty} \frac{e^{-u}}{u} \text{d}u
\end{equation}
with $\sigma_{ij} = (\sigma_i + \sigma_j)/2$.
The relaxation term in Eq.~\ref{Eq:MSA} is obtained from relaxation force acting
on each particle. In the linear response regime, this results in:
\begin{equation}
\frac{\delta F_j^{rel}}{F_j^{ext}} = -\frac{\kappa_{ij}^2}{3}
\frac{\sinh(\kappa_{ij} \sigma_{ij})}{\kappa_{ij} \sigma_{ij}} 
\int_{\sigma_{ij}}^\infty r h^0_{ij}(r)e^{-\kappa_{ij} r}\text{d}r
\end{equation}
with:
\begin{equation}
\kappa_{ij}^2 = 4\pi\ellB \frac{ \bar{\rho}_i q_i^2 D_i + \bar{\rho}_j q_j^2 D_j}{D_i+D_j} 
\end{equation}
For consistency with the hydrodynamic terms, the integral is considered up to
the same (second) order in the development of the equilibrium pair
distribution $h_{ij}^0$ within the MSA. This leads to
\begin{equation}
\frac{\delta F_j^{rel}}{F_j^{ext}} = \frac{\delta E^{(1)}}{E} + \frac{\delta
E^{(2)}}{E}
\end{equation}
where
\begin{eqnarray}
\frac{\delta E^{(1)}}{E} =
\frac{-\kappa_{ij}^2|q_iq_j|}{24\pi\varepsilon_0\varepsilon
\kB T\sigma_{ij}(1+\Gamma\sigma_i)(1+\Gamma\sigma_j)} \times \nonumber \\
\frac{1-e^{-2\kappa_{ij} \sigma_{ij}}}{\left[\kappa_{ij}^2 + 2\Gamma\kappa_{ij}
+2\Gamma^2 - 2\pi \ellB \sum \bar{\rho}_k\frac{q_k^2}{(1+\Gamma \sigma_k)}e^{-\kappa_{ij} \sigma_k}\right]} \nonumber
\end{eqnarray}
\begin{eqnarray}
\frac{\delta E^{(2)}}{E} = \frac{-\kappa_{ij} \ellB^2 q_i^2q_j^2\sinh(\kappa_{ij} \sigma_{ij})}
{ \sigma_{ij}(1+\Gamma\sigma_i)^2(1+\Gamma\sigma_j)^2} e^{2\kappa\sigma_{ij}} \nonumber \\
\times \int_{(\kappa_q+2\kappa)\sigma_{ij}}^\infty \frac{e^{-u}}{u}\text{d}u \nonumber
\end{eqnarray}
Within the MSA approximation, the only unkown parameters are the diameters of
the ions. The values reported in the main text are those of Ref.~\citenum{Turq2005}.

\begin{figure}
\includegraphics[width=8.5cm]{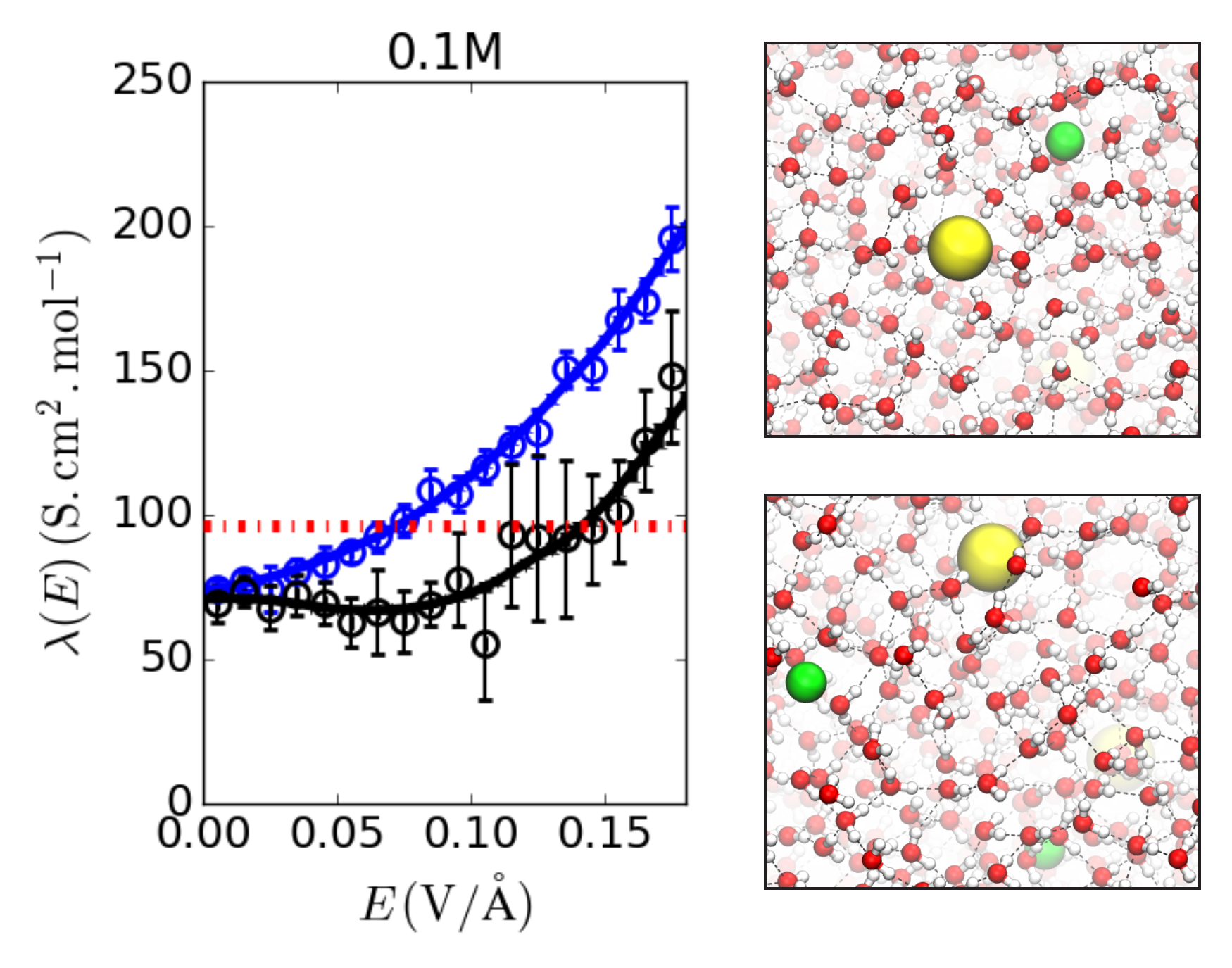}
\caption{(left) High field behavior of the lower concentration, explicit solvent system in an ensemble with the field on both the water and the ions, $E_{\mathrm{i}}=E_{\mathrm{w}}=E$ (black) or just on the ions $E_{\mathrm{i}}=E$ and $E_{\mathrm{w}}=0$ (blue).  Lines
are computed from reweighting $p_0(J, Q)$, and symbols are computed from finite differences of $\left<J\right>_E$ versus $E$. Errorbars are one standard deviation of the mean. The red dashed line corresponds to the Nernst-Einstein molar ionic conductivity $\lambda_{\mathrm{id}}$. (right) Charactoristic snapshots from the molecular dynamics simulations at $E=0$ (top) and $E=0.15 \mathrm{V/\AA}$ (bottom).  }
\label{Fig:HE}
\end{figure}
\section{High field behavior in the explicit model}
\label{Sec:HighE}
While the ionic conductivity under the moderate field considered in
Fig.~\ref{fig:Explicit} is weakly field-dependent, it does eventually grow
quadratically under very high fields as shown in Fig.~\ref{Fig:HE}. The
conductivity is found to exceed the Nernst-Einstein limit for free ions.
However, this behavior is unphysical. Under fields higher than $E=0.1~\mathrm{V/\AA}$, in our molecular simulations, we find that the dipoles of the water molecule all align with the high field, restraining dipole fluctuations, and leading to dielectric breakdown. Further, water molecules will start to spontaneously dissociate at such high fields,\cite{saitta2012ab} which is not allowed due to the constraints on water molecules in our model. 

\section*{References}

%

\end{document}